\documentclass[9pt,twocolumn,twoside]{osajnl}
\usepackage{siunitx}
\usepackage{textcomp}
\usepackage{titlesec}

\journal{optica} 
\setboolean{shortarticle}{true}

\title{Passive Laser Power Stabilization via an Optical Spring}

\author[1]{Torrey Cullen}
\author[1]{Scott Aronson}
\author[1]{Ron Pagano}
\author[2]{Marina Trad Nery}
\author[1]{Henry Cain}
\author[1*]{Jonathon Cripe}
\author[3]{Garret D. Cole}
\author[4]{Safura Sharifi}
\author[5]{Nancy Aggarwal}
\author[2]{Benno Willke}
\author[1]{Thomas Corbitt}

\affil[1]{Louisiana State University
Department of Physics \& Astronomy Baton Rouge, LA 70803}
\affil[2]{Max Planck Institute for Gravitational Physics (Albert Einstein Institute) and Institut f$\ddot{u}$r Gravitationsphysik, Leibniz Universit$\ddot{a}$t Hannover, Hannover, Germany}
\affil[3]{Crystalline Mirror Solutions, Santa Barbara, CA, USA}
\affil[4]{University of Illinois Urbana-Champaign Department of Physics Urbana, IL}
\affil[5]{Northwestern University Department of Physics and Astronomy, Evanston, IL}
\affil[*]{Current Affiliation: Laboratory for Physical Sciences, College Park, MD 20740, USA }

\begin{document}
\begin{abstract}
Metrology experiments can be limited by the noise produced by the laser involved via small fluctuations in the laser's power or frequency. Typically, active power stabilization schemes consisting of an in-loop sensor and a feedback control loop are employed. Those schemes are fundamentally limited by shot noise coupling at the in-loop sensor. In this letter we propose to use the optical spring effect to passively stabilize the classical power fluctuations of a laser beam. In a proof of principle experiment, we show that the relative power noise of the laser is stabilized from approximately $2 \times 10^{-5}$ Hz$^{-1/2}$ to a minimum value of $1.6 \times 10^{-7}$ Hz$^{-1/2}$, corresponding to the power noise reduction by a factor of $125$. The bandwidth at which stabilization occurs ranges from $400$ Hz to $100$  kHz. The work reported in this letter further paves the way for high power laser stability techniques which could be implemented in optomechanical experiments and in gravitational wave detectors.
\end{abstract}

\maketitle

\section{Introduction}
Laser power stabilization is important for many modern experiments, since power fluctuations can limit their sensitivity\cite{doi:10.1063/1.5040238,Matsumotothesis}. Currently, interferometric gravitational wave detectors require the most stringent power stability levels, where a relative power noise (RPN) of roughly $2 \times$ $10^{-9}$ Hz$^{-1/2}$ is required at $10$ Hz  by the Advanced LIGO detectors \cite{ALIGO}. A third generation of gravitational wave detectors is currently being planned, which will most likely require even higher power stability. So far strict requirements at low frequencies were mostly achieved using active power stabilization schemes, where an in-loop photodetector is used in conjunction with a feedback control loop. Those schemes are usually limited by noise sources coupling in the in-loop detector, and often require a large power detection which can exceed the power threshold of the in-loop sensors \cite{Marina_Thesis}. Recently, an alternative technique was demonstrated in which the full beam power of the laser and its fluctuations are sensed via a Michelson interferometer with a movable mirror \cite{marinapaper}. In this Letter, we propose to use a movable mirror in a Fabry-Perot cavity with a strong optical spring. The technique demonstrated in this paper is passive, and thus does not require a power sensor. We show here that this technique can provide large suppressions of classical power fluctuations such as to produce a beam in transmission of the cavity which is shot noise limited. In \cite{marinapaper}, a transfer and a sensing beam are used with a single movable cantilever mirror, in order to demonstrate active power stabilization from $1$ Hz to $10$ kHz. Here, we instead use a Fabry-Perot cavity with a strong optical spring to passively stabilize the power fluctuations transmitted by the cavity. Unlike other power stabilization techniques employing optical cavities \cite{Kwee:08}, the experiment proposed here provides power noise reduction below the cavity pole. This is an advantage, since it dispenses the use of long and high finesse cavities for stabilization at low frequencies.

\begin{figure}
\centering
\fbox{\includegraphics[width=\linewidth]{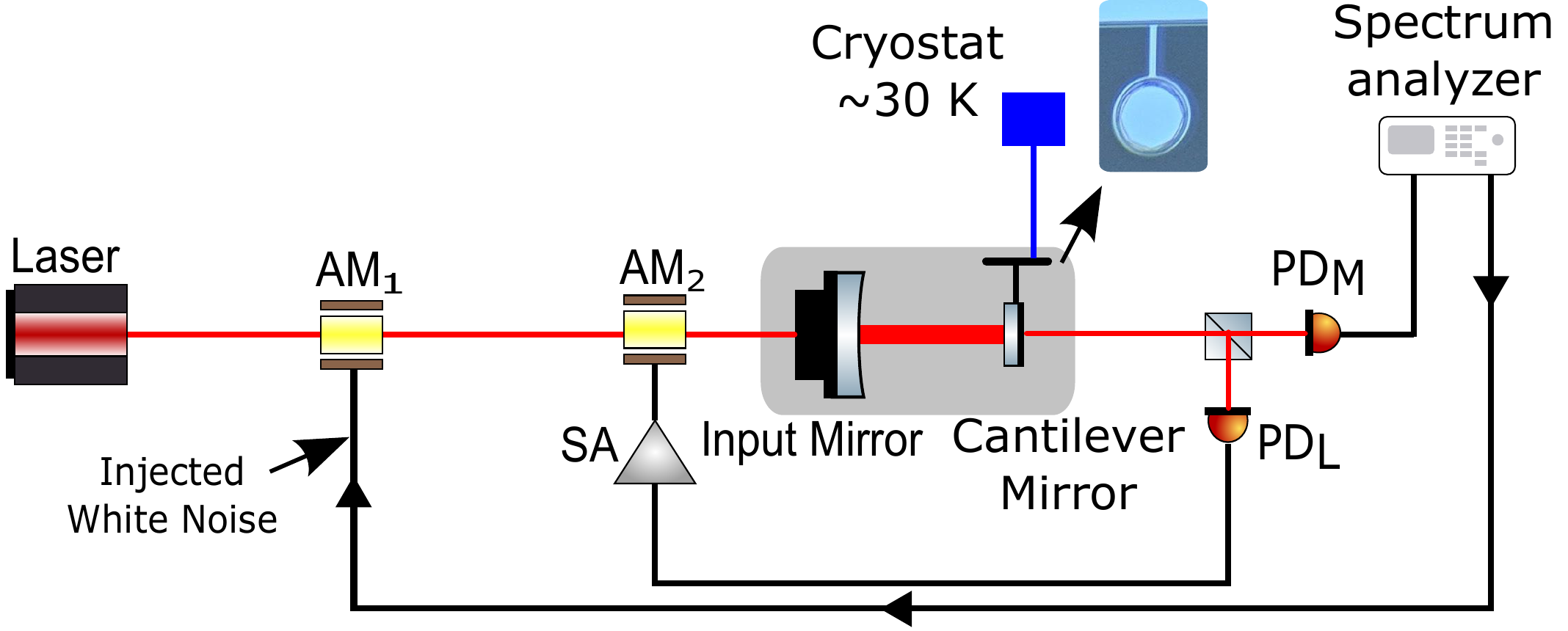}}
\caption{Simplified experimental setup of the passive laser power stabilization scheme via an optical spring. $\mathrm{PD_L}$ is used solely stabilize the optical spring above 100 kHz and not for active power stabilization.}
\label{fig:setup}
\end{figure}

We first start with a cavity comprised of a movable end mirror pumped with a laser (see Figure \ref{fig:setup}). By detuning the cavity away from its resonance, an optical spring \cite{cripe_thesis} is formed, whose dynamic response reduces power fluctuations in transmission of the cavity. To show that an optical spring can passively stabilize the power fluctuations of a laser we first start with the equation of motion of the movable mirror:
\begin{equation}
    m\ddot{x} = F_{rad}+F_{res}+F_{ext},
    \label{eq:2ndlaw}
\end{equation}
where $F_{rad}$ is the force on the mirror due to radiation pressure, $F_{res}$ is the restorative spring force, $F_{ext}$ is any external force, such as thermal noise, and \textit{m} is the mass of the mirror. Eq \ref{eq:2ndlaw} in the frequency domain then becomes:
\begin{equation}
    -m \Omega^2 \Tilde{x} = \frac{2\Tilde{P}_{\mathrm{circ}}}{c} - k \Tilde{x} + \Tilde{F}_{ext}.
    \label{eq:freqdomain}
\end{equation}
where \textit{c} is the speed of light, \textit{k} is the spring constant corresponding to the restorative mechanical spring force of the movable mirror, and $P_{\mathrm{circ}}$ is the circulating power in the cavity.

The power fluctuations ($\Delta P_{\mathrm{circ}}$) inside the cavity are dependent on the motion of the movable mirror and on the change of the maximally circulating power ($\Delta P_{\mathrm{max}})$ at resonance in the cavity, which in turn depends on the fluctuations of the injected laser power. These fluctuations can be written as:
\begin{equation}
    \Delta P_{\mathrm{circ}} = \frac{dP_{\mathrm{circ}}}{dx}\Delta x + \frac{dP_{\mathrm{circ}}}{dP_{\mathrm{max}}} \Delta P_{\mathrm{max}} .
    \label{eq:deltap}
\end{equation}
The circulating power in the cavity can be written in terms of the detuning ($\delta$, in units of linewidth) and the maximum power circulating in the cavity ($P_{\mathrm{max}}$) as \cite{cripe_thesis}:
\begin{equation}
    P_{\mathrm{circ}}(\delta) =  \frac{P_{\mathrm{max}}}{1+\delta^2}.
    \label{eq:p(delta)}
\end{equation}
Additionally,  the position of the movable mirror can be written in terms of the detuning as \cite{cripe_thesis}:
\begin{equation}
    x = \frac{\delta \lambda A}{8\pi},
    \label{eq:x}
\end{equation}
where $\lambda$ is the wavelength of light and $A$, the total losses of the mirrors, including their transmissivity, in which $A = 0$ is two perfectly reflective mirrors. Given the equations for power in terms of detuning (Eq \ref{eq:p(delta)}) and position in terms of detuning (Eq \ref{eq:x}), it is useful to rewrite Eq \ref{eq:deltap} as:
\begin{equation}
    \Delta P_{\mathrm{circ}} = \frac{dP_{\mathrm{circ}}}{d\delta}\frac{d\delta}{dx}\Delta x + \frac{dP_{\mathrm{circ}}}{dP_{\mathrm{max}}} \Delta P_{\mathrm{max}} .
    \label{longdeltap}
\end{equation}
Plugging in respective derivatives into Eq \ref{longdeltap} yields:
\begin{equation}
    \Delta P_{\mathrm{circ}} = -\frac{16 \pi P_{\mathrm{max}} \delta}{\lambda A (1+\delta^2)^2} \Delta x + \frac{\Delta P_{\mathrm{max}}}{1+\delta^2} .
    \label{eq:deltaplonger}
\end{equation}
%The ratio of the restorative force and the radiation pressure force in an optical spring arrangement is proportional to the square of the fundamental resonance frequency $\Omega_{\mathrm{fund}}$ over the optical spring frequency $\Omega_{os}$. For the experiment reported in this letter, we have:
%\begin{equation}
%    \left( \frac{\Omega_{\mathrm{fund}}}{\Omega_{os}} \right)^2 = %\left(\frac{0.9 \textrm{kHz}}{147 \textrm{kHz}}\right)^2 \approx 4 %\times 10^{-5}.
%\end{equation}
%This shows that the optical spring constant is much stronger than the mechanical spring constant and means the \textit{-kx} term can be neglected in Eq \ref{eq:freqdomain}, approximating the equation as:
%\begin{equation}
%    -m \Omega^2 \Delta \Tilde{x} \approx \frac{2 \Delta \Tilde{P}_{\mathrm{circ}}}{c}+\Delta \Tilde{F}_{ext}.
%    \label{eq:freqdomain2}
%\end{equation}
Solving Eq. \ref{eq:freqdomain} for $\Delta x$  to use in Eq \ref{eq:deltaplonger} yields:
\begin{equation}
    \Delta P_{\mathrm{circ}} = \frac{K_{os}\Delta P_{\mathrm{circ}}}{m(\Omega^2-\Omega^2_{\mathrm{fund}})}+\frac{\Delta P_{\mathrm{max}}}{1+\delta^2} +\frac{cK_{os} \Delta F_{ext}}{2 m(\Omega^2-\Omega^2_{\mathrm{fund}})},
    \label{eq:deltapalmost}
\end{equation}
where we have dropped the tilde notation and made the substitution for the optical spring constant $K_{os}$: \cite{Corbitt_thesis}
\begin{equation}
    K_{os} = \frac{32\pi P_{max}\delta}{\lambda A c(1+\delta^2)^2} .
\end{equation}
Note that this equation for the optical spring constant assumes that the response of the cavity is sufficiently slow such that it may be regarded as instantaneous.
Solving Eq \ref{eq:deltapalmost} for the intracavity power fluctuations yields,
\begin{equation}
\begin{aligned}
    \Delta P_{\mathrm{circ}} =& \frac{\Delta P_{\mathrm{max}}}{1+\delta^2}\left(\frac{\Omega^2-\Omega^2_{\mathrm{fund}}}{\Omega^2-\Omega^2_{os}-\Omega^2_{\mathrm{fund}}}\right)\\
    &-\frac{c}{2} \Delta F_{ext}\left(\frac{\Omega_{os}^2}{\Omega^2-\Omega^2_{os}-\Omega^2_{\mathrm{fund}}}\right),
    \label{eq:deltapfinal}
\end{aligned}
\end{equation}
where $\Omega_{\mathrm{fund}}$ is the resonance frequency of the fundamental mode of the cantilever mirror and $\Omega_{os}$ is the optical spring frequency, where both are defined as:
\begin{equation}
\begin{aligned}
    K_{os}= m  \Omega^2_{os}\hspace{.2cm}\mathrm{and}\hspace{.2cm}k= m  \Omega^2_{\mathrm{fund}}.
\end{aligned}
\end{equation}
An independent calculation of Eq \ref{eq:deltapfinal} can be found in \cite{nancy_thesis}.

For frequencies much smaller than the cavity pole, the connection between power fluctuations at the cavity input ($\Delta P_{\mathrm{in}}$) and the intracavity power fluctuations on resonance is $\Delta P_{\mathrm{max}} = PB*\Delta P_{\mathrm{in}}$, where $PB$ is the power buildup of the cavity. Hence, as seen from Eq \ref{eq:deltapfinal}, for frequencies much less than the optical spring resonance frequency ($\Omega  \ll \Omega_{os}$) and much greater than the fundamental mode of the cantilever mirror ($\Omega \gg \Omega_{\mathrm{fund}}$), the power fluctuations in the cavity are reduced compared to the injected field's power fluctuations by a factor of $(\frac{\Omega}{\Omega_{os}})^2$. We can also see that the external force, $F_{ext}$, imprints power fluctuations on the laser beam and will limit the stability achievable by this scheme. We have chosen an entirely classical derivation due to the fact that quantum radiation pressure effects are evaded when measuring the amplitude quadrature in transmission of the cavity \cite{PhysRevX.10.031065}.

The schematic of the experiment used to demonstrate this power fluctuation reduction is shown in Figure \ref{fig:setup}. It consists of an optomechanical cavity kept at cryogenic temperatures ($\sim 30$ K). The cavity is pumped with a $1064$ nm Nd:YAG nonplanar ring oscillator (NPRO) laser and is housed in a vacuum chamber kept at $10^{-8}$ torr. The movable mirror used in this setup is a cantilever mirror \cite{cripe_thesis} with a mass of $50$ nanograms, a fundamental frequency of $876$ Hz, and a quality factor of around $25400$. We note here that this quality factor is much larger than in previous experiments ($16000$) \cite{PhysRevX.10.031065,nancy_squeeze} due to the reduced temperature and pressure of the cryogenically cooled cavity. The input mirror in the cavity is a $0.5$ inch diameter, rigidly mounted mirror with a radius of curvature of $1$ cm. The cavity is just under $1$ cm long, with a pole greater than $100$kHz. Inside the vacuum chamber there is a vibration isolation platform which all the optics are mounted to, reducing seismic vibrations above $100$ Hz. 

\begin{figure*}
\centering
\footnotesize
\includegraphics[width=4.75in]{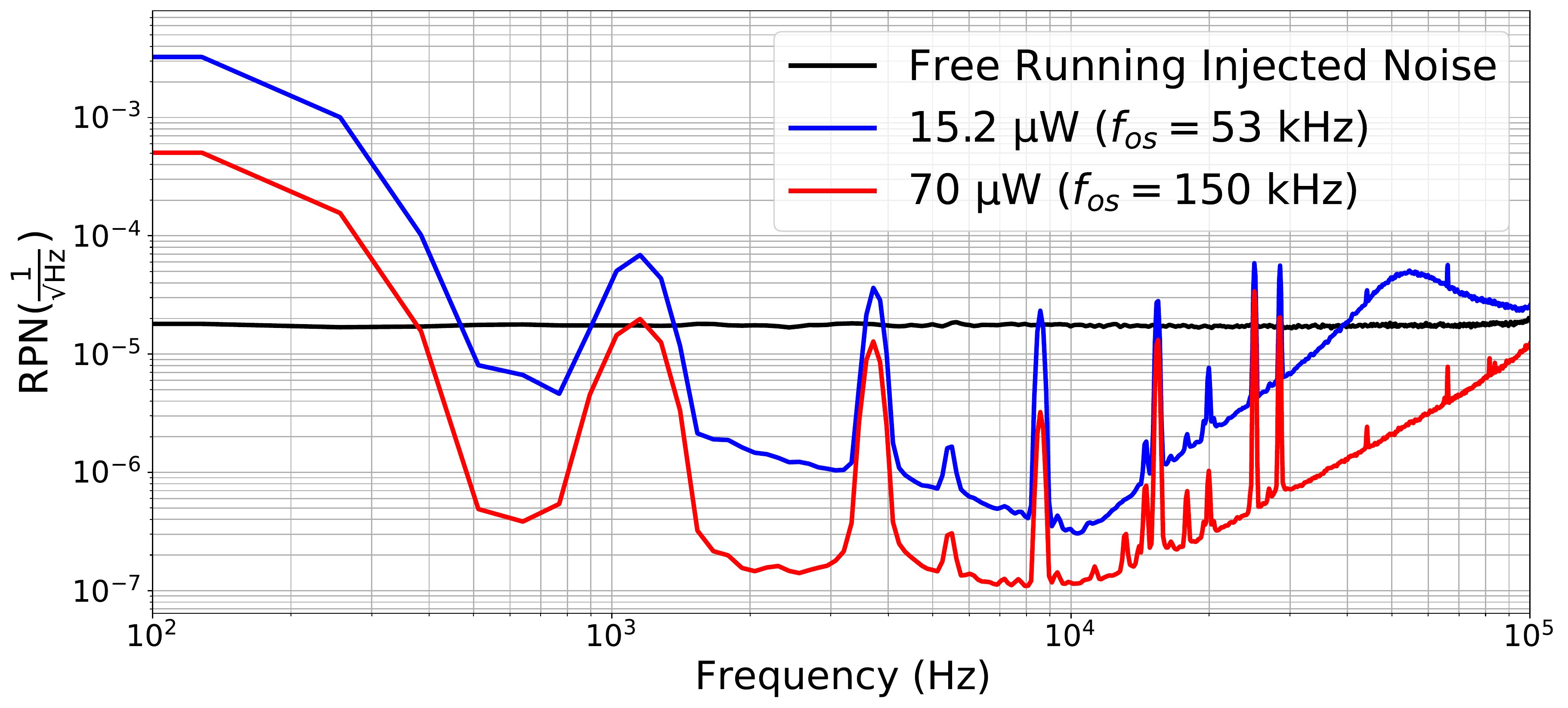}
\captionsetup{width=.91\linewidth}
\caption{Amplitude spectral density measurements showing reduction of classic power noise. White noise is added to the free running laser noise to bring the overall noise level up to approximately $2 \times 10^{-5}$ Hz$^{-1/2}$. The blue curve represents a cavity detuning of $3.6$ linewidths and the red curve a detuning $1.5$ linewidths. The large features at $3.7$ kHz, $15.5$ kHz and $28.5$ kHz are the coupling from the Yaw, Pitch, and Side-to-Side modes respectively of the cantilever mirror. }
\label{fig:result}
\end{figure*}
%\begin{figure*}
%  \begin{minipage}[c]{0.67\textwidth}
%    \includegraphics[width=\textwidth]{figures/Figure2.pdf}
%  \end{minipage}\hfill
%  \begin{minipage}[c]{0.3\textwidth}
%    \caption{Amplitude spectral density measurements showing reduction of classic power noise. White noise is added to the free running laser noise to bring the overall noise level up to approximately $2 \times 10^{-5}$ Hz$^{-1/2}$. The blue curve represents a cavity detuning of $3.6$ linewidths and the red curve a detuning $1.5$ linewidths. The large features at $3.7$ kHz, $15.5$ kHz and $28.5$ kHz are the coupling from the Yaw, Pitch, and Side-to-Side modes respectively of the cantilever mirror. } \label{fig:result}
%  \end{minipage}
%\end{figure*}
The requirement to keep a system like this stable is a positive optical spring constant and a positive damping coefficient \cite{Singh_PRL,17} which is not the case for our system. To keep the configuration of the cavity stable we employ a feedback loop actuating on an amplitude modulator ($\mathrm{AM}_2$ in Fig \ref{fig:setup}). This ensures the cavity stays at a constant detuning during a measurement. This feedback loop uses a photodetector ($\mathrm{PD_L}$ in Fig \ref{fig:setup}) in transmission of the cavity as the in-loop sensor and applies relevant correction signals to $\mathrm{AM_2}$ only for frequencies close to the optical spring frequency (i.e. above $100$ kHz for the red curve in Fig \ref{fig:result}). This feedback loop does not provide any power stabilization of the injected beam. Another photodetector in transmission of the cavity is used to monitor the power noise of the transmitted beam, labeled $\mathrm{PD_M}$.  Additional details describing the experimental setup can be found in \cite{nancy_squeeze,Cripe_QRPN}. 

In order to demonstrate the power stabilization we inject white noise into the amplitude modulator  before the cavity, $\mathrm{AM_1}$, at a voltage yielding a baseline relative power fluctuations of around $2\times10^{-5}$ Hz$^{-1/2}$. This is done to demonstrate that a large noise suppression is possible, since the free running relative power noise of the laser is at roughly $4\times10^{-6} \mathrm{Hz}^{-1/2}$ without the noise injection.
\vspace*{-.35 cm}
\section{Results}

\indent Figure \ref{fig:result} shows the results of the laser power stabilization for different optical spring strengths. The blue and red curves represent the cavity locked at a detuning corresponding to an optical spring frequency of $53 \pm 2$ kHz and $150 \pm 3$ kHz respectively. These optical spring frequencies correspond to a power transmitted by the cavity and incident on $\mathrm{PD_M}$ of $15.2$ and $70$ \SI{}{\micro \watt} respectively. Additionally, the input power for both these measurements is $6.5$mW. The black curve represents the sum of the free running laser noise and the intentionally imprinted white noise, which is measured at a photodetector just after $\mathrm{AM_1}$, not pictured in Fig \ref{fig:setup}. As seen from Fig \ref{fig:result}, the injected noise is suppressed by a greater amount with a stronger optical spring, as expected from Eq \ref{eq:deltapfinal}. The maximum suppression of the stronger optical spring measurement occurs at $7900$ Hz with a stabilized noise level of $1.6 \times 10^{-7}$ Hz$^{-1/2}$. This corresponds to the optical spring suppressing the injected noise by a factor of $125$. The blue curve measured for a cavity detuning of 3.6 linewidths has a steeper feature below $10$ kHz due to being thermal noise limited, whereas the red curve (1.5 linewidths detuned) is mainly shot noise limited. Above $10$kHz, both curves are limited by the noise suppression provided by the optical spring.
%\begin{figure}[h]
%\centering
%\footnotesize
%\includegraphics[width=\linewidth]{figures/Figure2.pdf}
%\caption{Amplitude spectral density measurements showing reduction of classic power noise. White noise is added to the free running laser noise to bring the overall noise level up to approximately $2 \times 10^{-5}$ Hz$^{-1/2}$. The blue curve represents a cavity detuning of $3.6$ linewidths and the red curve a detuning $1.5$ linewidths. The large features at $3.7$ kHz, $15.5$ kHz and $28.5$ kHz are the coupling from the Yaw, Pitch, and Side-to-Side modes respectively of the cantilever mirror. }
%\label{fig:result}
%\end{figure}

\begin{figure*}[htp]
\centering
\includegraphics[width =5.4in]{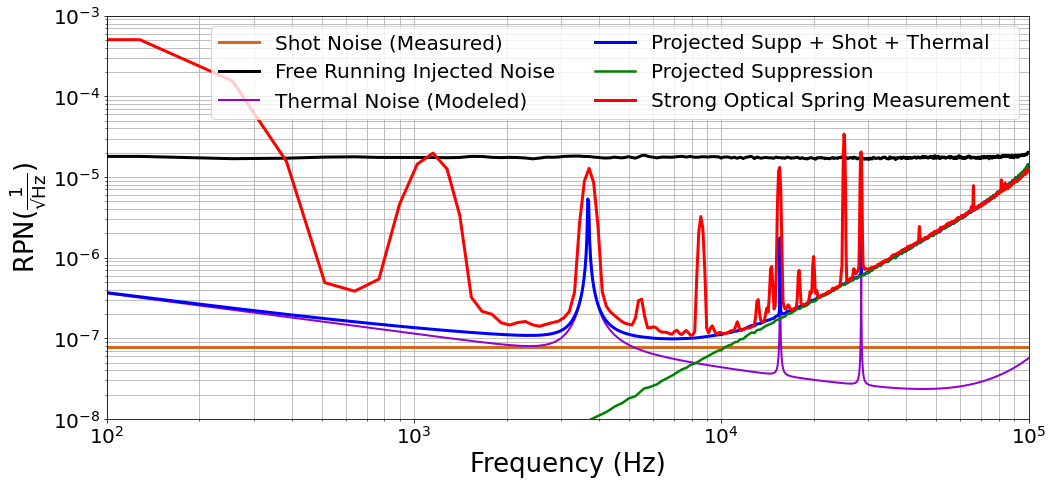}
\captionsetup{width=.9\linewidth}
\caption{\SI{70}{\micro \watt} power noise spectrum post cavity measurement with its associated fundamental noise limitations, which includes relative shot noise, the calculated (modeled) RPN, and projected suppression of the cavity. The circulating power corresponding to \SI{70}{\micro \watt} of transmitted power is $250$ mW.}
\label{fig:resultbudget}
\end{figure*}

\indent Figure \ref{fig:resultbudget} compares the highest power noise suppression measurement (red curve) with an uncorrelated sum of fundamental limits of this experiment (blue curve) and the power noise suppression by the optical spring. These fundamental limits are comprised of relative shot noise and thermal noise of the cantilever mirror added in quadrature. The total limit for the relative power noise detected by $\mathrm{PD_M}$ can be obtained by dividing Eq. \ref{eq:deltapfinal} by the mean circulating power $\mathrm{P_{circ}}$, since, below the cavity pole, the classical RPN transmitted by the cavity ($\mathrm{RPN_M}$) should be the same as the classical RPN inside the cavity. Hence, the amplitude spectral density (ASD) of the RPN at $\mathrm{PD_M}$ is:
\begin{equation}
\begin{aligned}
    \mathrm{RPN_M}^2&=  \mathrm{RPN_{in}}^2\left(\frac{\Omega^2-\Omega^2_{\mathrm{fund}}}{\Omega^2-\Omega^2_{os}-\Omega^2_{\mathrm{fund}}}\right)^2 \\
    &-\left(\frac{c F_{\mathrm{ext, ASD}}}{2 P_{circ}}\right)^2\left(\frac{\Omega_{os}^2}{\Omega^2-\Omega^2_{os}-\Omega^2_{\mathrm{fund}}}\right)^2+ \frac{2hc}{\lambda P_M} \mathrm{,}
    \label{eq:fullrpn}
\end{aligned}
\end{equation}
where $\mathrm{RPN_{in}}$ and $\mathrm{F_{ext,ASD}}$ are the ASD of the RPN at the input of the cavity, and of the external force.
%\begin{figure*}
%  \begin{minipage}[c]{0.75\textwidth}
%    \includegraphics[width=\textwidth]{figures/Figure3_Full_Budget.png}
%  \end{minipage}\hfill
%  \begin{minipage}[c]{0.25\textwidth}
%    \caption{\SI{70}{\micro \watt} power noise spectrum post cavity measurement with its associated fundamental noise limitations, which includes relative shot noise, the calculated (modeled) RPN, and projected suppression of the cavity. The circulating power corresponding to \SI{70}{\micro \watt} of transmitted power is $250$ mW.} \label{fig:resultbudget}
%  \end{minipage}
%\end{figure*}
The first term in Eq \ref{eq:fullrpn} represents the power noise suppression by the optical spring. The third term in Eq \ref{eq:fullrpn} is the relative shot noise of the measured power on $\mathrm{PD_M}$, where \textit{h} is Planck's constant, and $P_M$ is the mean power transmitted by the cavity and detected on $\mathrm{PD_M}$. This relative shot noise level is additionally corroborated by measuring the shot noise of the photodetector experimentally. Both by experimental methods and the relative shot noise term in Eq \ref{eq:fullrpn} the average shot noise level was determined to be $7.3 \times 10^{-8}$ Hz$^{-1/2}$ and is depicted by the brown curve in Fig \ref{fig:resultbudget}. The next part in the fundamental limits curve is the thermal noise term contribution:
\begin{equation}
   \mathrm{RPN_{tn}} = \left(\frac{\Omega^2-\Omega^2_{\mathrm{fund}}}{\Omega^2-\Omega^2_{os}-\Omega^2_{\mathrm{fund}}}\right) \frac{c \Omega_{\mathrm{fund}}}{ P_\mathrm{circ}}\sqrt{\frac{ k_b T m}{Q \Omega}},
\end{equation}
where $\Omega_{\mathrm{fund}}$ is the fundamental resonance frequency of the cantilever mirror, \textit{Q} is the structural quality factor of the movable mirror, \textit{T} is the temperature, and $k_b$ is the Boltzmann constant. This equation was calculated by substituting $F_\mathrm{ext}$ in Eq \ref{eq:fullrpn} by the thermal noise force considering structural damping \cite{Saulson} and is the minimum relative power noise in transmission of the cavity, limited by thermal noise of the movable mirror. This quantity is a similar result as obtained in \cite{marinapaper,marinapaper2}, but here with a dependence on the intracavity power. This is an advantage for this scheme since a trade off of using a mirror with high spring constant can be made by increasing the intracavity power. 

Because the laser beam is not perfectly centered on the movable mirror, we see a coupling of the cantilever modes, pitch ($3.7$kHz), yaw ($15.5$kHz), and side-to-side ($28.5$ kHz), in our measurement. If the beam was perfectly aligned these features would not be observed. In order to account for this in our fundamental limits we use a modeling code that uses the two photon formalism \cite{Corbitt_mathematical} that accounts for the centering of the beam when calculating thermal noise. The result for the thermal noise model in this experiment is shown by the lilac curve in Figure \ref{fig:resultbudget}. The temperature recorded for these measurements refers to the upper limit of the temperature of the cavity. This is because the cryostat introduces mechanical vibrations strong enough to interfere with the locking capabilities of the cavity. For this reason, the cryostat compressor is turned off and the cavity slowly warms as the measurement is performed. Generally, by the time a measurement is finished the cavity is at ~$30$K. \\
\indent The final contribution to the total limit is the residual input noise limited by the suppression by the optical spring. This is calculated by taking the injected noise and multiplying by the suppression term (first term) in Eq \ref{eq:fullrpn}. For the parameters of the stronger optical spring measurement, the optical spring has the potential to provide a power noise suppression  of approximately $3 \times 10^4$ at $10$ Hz. This factor is quite large and thermal or quantum noise typically limit the performance at low frequency. \\
\indent Given these parameters, we find the total noise budget agrees with the measured spectrum for most frequencies. At low frequencies, the experiment is limited by seismic noise, hence the additional noise in the measurement with respect to the blue curve in Figure \ref{fig:resultbudget}. In theory, it is possible to lock a Fabry-Perot cavity like this one without the use of feedback, instead using a large, positively detuned carrier beam and a small, negatively detuned sub-carrier beam \cite{Singh_PRL,17}. This has been tested to show the stability of the double optical spring effect, but not yet on its ability to stabilize the power of the laser. In this regime however, it would be possible to lock the cavity and have the power stabilized without the use of feedback anywhere. \\
\indent In order to achieve a lower RPN and a larger power in transmission of the cavity, the input power, and therefore the circulating power, needs to be increased. This reduces the contribution from thermal noise and shot noise, as shown in Eq. \ref{eq:fullrpn}. In this experiment, the circulating power of $250$ mW was limited by the damage threshold of the micro-oscillators. However, the suspended mirrors described in \cite{PhysRevA.92.033825}, as well as \cite{PhysRevLett.124.221102}, utilize parameters that could provide a total RPN close to the needs of current gravitational wave detectors. With a 5-mg suspended mirror having a resonance on the order of $10$ Hz, Q of $1 \times 10^5$, an intracavity power of $200$ W, and a temperature of $20$K, we find that the RPN for such a system would be $1.4 \times 10^{-9} \mathrm{Hz}^{-1/2}$ at 10 Hz and $4.4 \times 10^{-10} \mathrm{Hz}^{-1/2}$ at 100 Hz. These parameters are realizable and would give the RPN levels necessary for aLIGO.

\subsection*{Funding}
\small{\noindent T. Cullen, SA, RP, HC, and T. Corbitt are supported by the NSF grants PHY-2110455 and PHY-1806634. MTN and BW are funded by the Deutsche Forschungsgemeinschaft (DFG, German Research Foundation) under Germany’s Excellence Strategy – EXC-2123 QuantumFrontiers – 390837967.}
\subsection*{Disclosures}
\small{The authors declare no conflicts of interest.}
\bibliography{main.bib}
%\bibliographyfullrefs{bib.bib}

\noindent© 2022 Optica Publishing Group. Users may use, reuse, and build upon the article, or use the article for text or data mining, so long as such uses are for non-commercial purposes and appropriate attribution is maintained. All other rights are reserved. In Preparation.
\end{document}